\documentclass[preprint,pre,aps,showpacs]{revtex4}
\usepackage{graphicx}
\begin{document}

\title{Fluctuations in the number of intermediate mass fragments in small projectile
like fragments.}

\author{S. Mallik, G. Chaudhuri}
\affiliation{Variable Energy Cyclotron Centre, 1/AF Bidhannagar,
Kolkata 700064, India}
\author{S. Das Gupta}
\affiliation{Physics Department, McGill University,
Montr{\'e}al, Canada H3A 2T8}

\date{\today}

\begin{abstract}
The origin of fluctuations in the average number of intermediate mass fragments
seen in experiments in small projectile like fragments is discussed.  We
argue that these can be explained on the basis of a recently proposed
model of projectile fragmentation.
\end{abstract}


\maketitle
This report is an outgrowth of a recent work \cite{Mallik2} where we introduced a model
of projectile fragmentation in heavy ion collisions.  The model has three
ansatzs: at a given impact parameter a certain part of the projectile is
sheared off forming a projectile like fragment (PLF) with charge number $Z_s$
and neutron number $N_s$.  This is called abrasion.  This hot PLF expands to
about one-third the normal density and then breaks up into several composites
at a given temperature $T$.  This break up is according to a canonical
thermodynamical model (CTM).  The composites are hot and will decay by
evaporation.  For details, please see \cite{Mallik2}.  The main contention of reference
\cite{Mallik2} was that the temperature $T$ must be taken to be a function of the
impact parameter $b$.

Results of the following experiment done at the SIS heavy-ion synchrotron
at GSI Darmstadt are published \cite{Ogul}.  In an event let us denote the number of
intermediate mass fragments (IMF) (charge $z$ between 3 and 20) by $N_{IMF}$.
In the same event denote by $Z_{bound}$=sum of all the charges in the
PLF minus the charges of $z=1$ particles (proton, deuteron and triton).
After many events one can plot $M_{IMF}$ $\it {vs}$ $Z_{bound}$ where $M_{IMF}$
is the average of $N_{IMF}$.  The data and comparison with the theoretical
calculation done in \cite{Mallik2} is shown in Fig. 1.\\
\begin{table}[h]
\begin{center}
\begin{tabular}{|c|c|c|c|}
\hline
&\multicolumn{3}{c|}{$M_{IMF}$}\\
\cline{2-4}
$Z_{bound}$& $Sn^{107}$ & $Sn^{124}$ & $La^{124}$\\
\hline
3 & 1.000 & 1.000 & 1.000\\
4 & 0.140 & 0.000 & 0.178\\
5 & 1.000 & 1.000 & 1.000\\
6 & 0.430 & 0.565 & 0.620\\
7 & 1.062 & 1.078 & 1.092\\
\hline
\end{tabular}
\end{center}
\caption{ Experimentally measured $M_{IMF}$ at small $Z_{bound}$ values for
 $Sn^{107}$ on $Sn^{119}$, $Sn^{124}$ on $Sn^{119}$ and $La^{124}$ on $Sn^{119}$ reactions. }
\end{table}
The overall feature of the figure is that the general shapes of the
theoretical and experimental curves agree.  The $b$ dependence of $T$ is
crucial for this (as explained in \cite{Mallik2}).  However, there are
significant fluctuations in the experimental values of $M_{IMF}$ for
low values of $Z_{bound}$ whereas theory completely misses these fluctuations.
In this note we explain (a)why these fluctuations arise, (b) how staying
within the main ingredients of the theoretical model but using more realistic
parameters we can reproduce the fluctuations and (c) why the calculation in
\cite{Mallik2} missed the fluctuations seen in small PLF's.\\

First we explain how the fluctuations arise.
We have $Z_{bound}$=$Z_s$ minus the sum of charges of all $z$=1 particles
(protons, deuterons and tritons).  Also a particle is considered to be an
IMF if its charge $z$ is between 3 and 20.  Just these two conditions and
some general knowledge of low-mass nuclei allow us to reach some interesting
conclusions.

If $Z_{bound}$=3 it guarantees that we have a $^A_3$Li nucleus.  Thus for
$Z_{bound}$=3 $M_{IMF}$=1.
If $^A_3$Li decays by a proton emission we are no longer in
$Z_{bound}$=3 but degenerate into $Z_{bound}$=2.  Also there is no IMF.
If it decays by neutron
emission to a particle stable state of a different isotope of Li, we still
have $M_{IMF}$=1.
There are several particle stable states of Li so $Z_{bound}=3$, $M_{IMF}$=1
is always satisfied.

Let us consider now $Z_{bound}$=4.  For $Z_{bound}$=4 we can
have a Be nucleus with $N_{IMF}$=1 but it can also decay into two He isotopes
which still retains $Z_{bound}$=4 but with $N_{IMF}$=0.
We therefore expect to have $Z_{bound}$=4 and
$M_{IMF}=X$ where $X$ is less than 1.  But unfortunately $X$ is not the same
value for all Be nuclei.  We will soon demonstrate how $X$ could be
determined for each Be nucleus but the fact that $X$ varies from one isotope
of Be to another isotope of Be makes the evaluation of $M_{IMF}$
in the case of $Z_{bound}$=4 a lengthy procedure.

If $Z_{bound}$=5 we have either one Boron nucleus or a Li nucleus plus a He
nucleus.  In both the cases $M_{IMF}$=1.
If the Boron nucleus sheds a proton, the
status drops to $Z_{bound}$=4 and we are back to the $Z_{bound}$=4 case.
If the Boron nucleus sheds one or more
neutrons to reach a particle stable state we maintain $Z_{bound}=5, M_{IMF}$=1.
If Boron decays into a Li and He two things can happen.  We reach a particle
stable state of Li and we have $Z_{bound}=5$, $M_{IMF}$=1.  If the Li sheds a
proton we no longer have $Z_{bound}$=5.  Thus so long as we have $Z_{bound}$=5
we have $M_{IMF}$=1.

We want to get back to the case of $Z_{bound}$=4.  Now we need to bring in
details of the model.  Two modifications are made.  To carry out CTM one
needs to put in the partition function of each composite into which the
hot abraded PLF can break into.  In our previous
calculation, except for nuclei upto $^4$He, we used the liquid-drop model
for the ground state energy and the Fermi-gas model for excited states.
For small PLF's this is inaccurate and we put in experimental values of ground
state and excited state energies.  Usually all excited states upto 7.5 MeV
are included.  Next we consider the decays of hot composites resulting from
CTM.  Previously we used an evaporation code.  We replace this by actual
decay data whenever possible.  In practical terms this means the
following.  A nucleus has many energy levels and a hot nucleus means that
the probability of occupation of a state $i$ is proportional to
$s_iexp(-exc(i)/T))$ where $s_i$ is the spin degeneracy and $exc(i)$ is the
excitation energy.  The decay of the state $i$ is taken from data table
\cite{Ajzenberg1, Ajzenberg2, Tilley1, Tilley2, www1}
where available or guessed from systematics.
We take $T$ to be 7 MeV suggested by our past work \cite{Mallik2}.

It is useful to list first the deacy properties of hot Be nuclei. These are
computed at $T$=7 MeV.

$^6_4$Be: This decays into \cite{Ajzenberg1} $^4_2$He
plus 2 protons so this counts as $Z_{bound}$=2 and $M_{IMF}$=0.

$^7_4$Be: The lowest 2 states are particle stable.  Population into any of
these gives $Z_{bound}$=4 and one IMF.  The probability
of this occurring is 0.406.  The other states decay into $^4_2$He plus
$^3_2$He leading to $Z_{bound}$=4 and $N_{IMF}$=0. Thus for $^7_4$Be
we have $Z_{bound}$=4 and $M_{IMF}$=0.406.

$^8_4$Be: this occurs as resonances of two $^4_2$He
so here $Z_{bound}$=4 and $M_{IMF}$=0.

$^9_4$Be:  Only the ground state is particle stable, the rest decay to
neutron plus two alphas.  The occupation probability in the ground state
is 0.193. So here $Z_{bound}$=4 and $M_{IMF}$=0.193.

$^{10}_4$Be: Here we have taken all the levels upto 6.26 MeV
(summed occupation probability=0.604) to give
$Z_{bound}$=4 and 1 IMF and rest of the levels upto 9.3 MeV to give
$Z_{bound}$=4 and 0 IMF.  Thus $Z_{bound}$=4 and $M_{IMF}$=0.604.

$^{11}_4$Be: Here the lowest two levels have $Z_{bound}$=4 and $N_{IMF}$=1
and the probability of occupation 0.1567.  The higher levels, with
summed occupation probability 0.8433 go to $^{10}_4$Be+n.
We have assigned them
$Z_{bound}$=4 and $M_{IMF}$=0.604.  Thus we take $^{11}_4$Be to give
$Z_{bound}$=4 and $M_{IMF}$=0.666.

Let us now outline how we calculate $M_{IMF}$ for $Z_{bound}$=4
for collisions of $^{107}$Sn, $^{124}$Sn and $^{124}$La on $^{119}$Sn.
Although our discussion will be limited to $Z_{bound}$=4, the method
can be extended to higher values of $Z_{bound}$ except that the complexity
increases very rapidly.   The method of obtaining the
abrasion cross-section for a PLF with given $Z_s,N_s$ is given in \cite{Mallik2}.
For $Z_{bound}$=4 we need to consider $Z_s$=4 (most important) and higher.
Once a PLF with given $Z_s,N_s$ is formed it will expand to one-third the
normal nuclear density and break up into hot composites.  Just as we could
characterize a hot Be nucleus by a $Z_{bound}$ and $M_{IMF}$ we can ascribe
to each $Z_s,N_s$ a probability of obtaining $Z_{bound}$=4 with an associated
$M_{IMF}$. (An example below shows how this can be done.)  Table II compiles
these values (last two columns).
\begin{table}[h]
\begin{center}
\begin{tabular}{|c|c|c|c|c|c|c|}
\hline
& &\multicolumn{3}{c|}{Cross-section (mb)}& &\\
\cline{3-5}
$Z_s$ & $N_s$ & $Sn^{107}$ & $Sn^{124}$ & $La^{124}$ & $P(Z_{bound}=4)$ & $M_{IMF}$ \\
\hline
4 & 3 & 0.6597 & 0.0 & 0.0 & 0.605 & 0.406\\
4 & 4 & 8.9445 & 0.5644 & 5.1102 & 0.583 & 0.043\\
4 & 5 & 10.5290 & 4.5058 & 8.3227 & 0.569 & 0.165\\
4 & 6 & 1.1099 & 8.8110 & 0.6300 & 0.486 & 0.448\\
4 & 7 & 0.0 & 3.7233 & 0.0 & 0.467 & 0.592\\
\hline
5 & 4 & 0.4406 & 0.0 & 0.0 & 0.6087 & 0.074\\
5 & 5 & 7.7417 & 0.0 & 4.5875 & 0.2842 & 0.125\\
5 & 6 & 12.8264 & 2.1752 & 11.4830 & 0.2304 & 0.194\\
5 & 7 & 1.5797 & 9.2460 & 2.5002 & 0.1926 & 0.336\\
5 & 8 &0.0 & 8.5111 & 0.0 & 0.1782 & 0.498\\
\hline
\end{tabular}
\end{center}
\caption{ Abrasion cross-sections (in millibarns) for a given ($Z_s$, $N_s$) for $Sn^{107}$, $Sn^{124}$ and $La^{124}$ on $Sn^{119}$. $P(Z_{bound}=4)$ gives the probability of obtaining $Z_{bound}=4$ for a given $Z_s$, $N_s$ and $M_{IMF}$ is the corresponding average multiplicity of intermediate mass fragments.}

\end{table}\\
Utilizing also the values of the
abrasion cross-sections for $Z_s,N_s$ for the three reactions
(also given in the Table) we get the
desired results.  For $Z_{bound}$=4, $M_{IMF}$=0.145(0.14) for $^{107}$Sn
beam, 0.151(0.178) for $^{124}$La beam and 0.38(0) for $^{124}$Sn beam.  The
experimental values are enclosed by parenthesis.  Except for $^{124}$Sn beam
our results approximately correspond to the experimental data.  The value 0 for
$^{124}$Sn is a mystery.  In any model we can think of the result should
not be 0 or very different from the other two.  In any case we have reproduced
the fluctuation: $M_{IMF}$ drops from 1 at $Z_{bound}$=3 to much lower value
at $Z_{bound}$=4 and back again to 1 at $Z_{bound}$=5.  It is very long to do
a quantitative estimate for $Z_{bound}$=6.  This will arise from $Z_s$=6 and
higher.  A study of the CTM results of $Z_s=6,N_s=7$ shows the following.
There is a significant probability of reaching a Carbon
nucleus ($Z_{bound}$=6).  This will produce a $M_{IMF}\approx$1.  There is a comparable
probability of obtaining $Z_{bound}$=6 with a $^8$Be nucleus (zero IMF) and
another He nucleus and also a $^9$Be nucleus ($M_{IMF}$=0.193) and another
He nucleus.  The probability of reaching two Li nuclei post CTM is
non-negligible but the chances of any one or both of them decaying by
alpha or proton emission (thereby dropping below $Z_{bound}$=6) are quite
high (0.88).  A theoretical value for $M_{IMF}\approx$0.5 seems quite plausible.
It is the fragility of $Be$ nuclei which produces the dip in $M_{IMF}$ for $Z_{bound}$=4
and is also responsible for the dip at $Z_{bound}$=6.
\\
\begin{table}[h]
\begin{center}
\begin{tabular}{|c|c|c||c|c|c|}
\hline
$A$ & $Z$ & $<n_{A,Z}>$ &$A$ & $Z$ & $<n_{A,Z}>$ \\
\hline
9 & 4 & $4.7518\times10^{-1}$ & 5 & 3 & $8.0475\times10^{-3}$\\
8 & 4 & $8.8281\times10^{-2}$ & 5 & 2 & $1.2987\times10^{-1}$\\
8 & 3 & $3.8108\times10^{-2}$ & 4 & 2 & $1.3924\times10^{-1}$\\
7 & 4 & $5.8888\times10^{-3}$ & 3 & 2 & $2.4670\times10^{-2}$\\
7 & 3 & $1.0194\times10^{-1}$ & 3 & 1 & $1.0669\times10^{-1}$\\
6 & 4 & $1.1442\times10^{-4}$ & 2 & 1 & $1.5234\times10^{-1}$\\
6 & 3 & $1.0470\times10^{-1}$ & 1 & 1 & $7.4883\times10^{-2}$\\
6 & 2 & $2.1149\times10^{-2}$ & 1 & 0 & $1.8152\times10^{-1}$\\
\hline
\end{tabular}
\end{center}
\caption{ Multiplicity of different fragments produced by CTM from the abraded nucleus $Z_s=4$, $N_s=5$ at $T=7.0$ MeV.}
\end{table}
As promised, let us give an example how for a given $Z_s,N_s$ the
probability of occurrence
of $Z_{bound}$=4 and the associated $M_{IMF}$ can be computed (last two columns of Table II).  Consider $Z_s=4,N_s=5$.
To start with, the average numbers of each composite resulting from the
CTM break up of $Z_s=4,N_s=5$ system are listed in Table III.
But in a simple case like this, this  can also give, with little
effort, the
probability of a channel or the probability of a sum of channels.
From the table,
the average number of $^9_4$Be is $\approx$0.475.  This is a channel where
only $^9_4$Be and nothing else appears.  Thus there is a probability of
0.475 of reaching $Z_{bound}$=4 and $M_{IMF}$=0.193.
Next, looking at the table,
the average number of $^8$Be is
$\approx$0.088.  This comes from a channel where there is one $^8$Be
and one neutron.  So we have a probability of 0.088 of reaching $Z_{bound}$=4
with $M_{IMF}$=0.0.
Next from the table, the average number of $^8_3$Li
is $\approx$0.038.  This has to occur in combination with
a proton.  Clearly, this is channel with $Z_{bound}$=3, so this does not
concern us presently.  Next,
from the table, the average number of $^7_4$Be is $\approx$0.006.  This
is a channel which has one $^7_4$Be and 2 neutrons.
Thus we have a probability of 0.006 of reaching $Z_{bound}$=4 with $M_{IMF}$=
0.406.  We have exhausted all the channels for reaching $Z_{bound}$=4.
Summing up with appropriate
weightage, from $Z_s=4,N_s=5$ the probability of reaching $Z_{bound}$=4
is 0.569 with $M_{IMF}$=0.165.

We can repeat similar arguments for other $Z_s,N_s$ in Table II.  The cases
of $Z_s$=5 are more complicated.

We now try to answer why the calculation of \cite{Mallik2} failed to produce any
fluctuation.  There are many reasons (use of liquid-drop model and
non-recognition of the fragility of Be nucleus etc.) but the most
interesting reason is different.

The prescription we used for $M_{IMF}$ vs. $Z_{bound}$ is the following.
At a given $b$, abrasion gives an integral $Z_s$ (and an integral $N_s$).
This system expands, then dissociates by CTM and the hot composites which
are the end results of the CTM, can evaporate light particles to give the
final distribution.  From this we obtained $M_{IMF}$ and we considered
$Z_{bound}$ to be given by $Z_s-\sum_in_{z=1}(i)$ where $n_{z=1}(i)$ stands for
the average multiplicity of proton/deuteron/triton.  This prescription
does not match exactly the experimental procedure.  Experimentally $Z_{bound}$
is obtained event by event and in every event $Z_{bound}$ is an integer
(sum of all charges from PLF minus number of particles with $z$=1).  From
many events with the same $Z_{bound}$ one can obtain $M_{IMF}$.  In our
calculations although $Z_s$ is an integer $Z_s-\sum_in_{z=1}(i)$ will
usually be non-integer since the $n_{z=1}(i)$'s (average number of composite
$i$) are.

Our calculation can map much better into a different experiment.
In this experiment $Z_s$ is measured but $z$=1 particles are not
subtracted.  One then obtains $M_{IMF}$ for each $Z_s$.  This problem
is simpler:  given a total number of particles, what is
$M_{IMF}$?  But in the reported experiment one asks a more exclusive
question : when the particles
are fractured in a certain way (a given number of particles with charge
greater than 1) what is $M_{IMF}$?  In our prescription we get a
non-integral value for $Z_{bound}$ and what we are obtaining is an
average of $M_{IMF}$ done over $M_{IMF}$ belonging to different but
neighbouring values of integral $Z_{bound}$.  This would be quite wrong
if values of $M_{IMF}$ belonging to neighbouring $Z_{bound}$'s differ
strongly (as it happens for very small systems) but for large systems
the difference would be small and our prescription is adequate for
an estimate.\\
This work was supported in part by Natural Sciences and Engineering Research
Council of Canada. The authors are thankful to Prof. Wolfgang Trautmann for access to experimental data. S. Mallik is thankful for a very productive and enjoyable stay at McGill University for this work. S. Das Gupta wants to thank Prof. Jean Barrette for discussions.

\begin{figure}
\includegraphics[height=3.25in,width=5.25in]{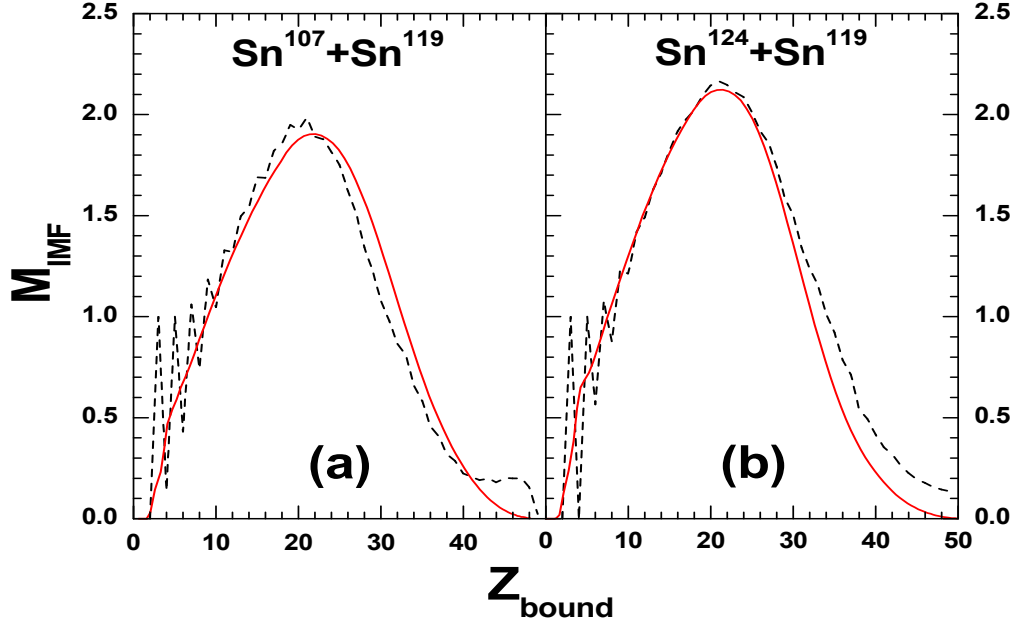}
\label{fig3}
\caption{ (Color Online) Mean multiplicity of intermediate-mass fragments $M_{IMF}$, as a function of $Z_{bound}$ for (a) $^{107}$Sn on $^{119}$Sn and (b) $^{124}$Sn on $^{119}$Sn reaction (red solid lines). Temperature is impact parameter ($b$) dependent and falls off linearly with b from $7.5$ MeV at $b$=0 to $3$ MeV at maximum value of $b$. The experimental results \cite{Ogul} are shown by the black dashed lines. }
\end{figure}

\end{document}